\pdfoutput=1

\documentclass[twocolumn,lscape,prl,aps,10pt,nofootinbib,showpacs,superscriptaddress,preprintnumbers]{revtex4-1}

\newcommand\sect[1]{{\it #1.}---}
\usepackage[utf8]{inputenc}
\usepackage{graphicx, color}

\usepackage[svgnames]{xcolor}
\usepackage[colorlinks,citecolor=DarkGreen,linkcolor=FireBrick,linktocpage]{hyperref}

\usepackage{subfigure}
\usepackage{dcolumn}
\usepackage{bm}
\usepackage{amsmath}
\usepackage{amsthm}
\usepackage{ascmac}
\usepackage{xparse} 
\usepackage{mathtools} 
\usepackage{blkarray} 
\usepackage{amscd,verbatim}
\usepackage{amssymb}
\usepackage[all]{xy}
\usepackage{tikz}
\usepackage{tikz-cd}
\usepackage{pgf,pgfplots}
\usepackage{enumitem}
\usepackage{caption}
\usepackage{subcaption}
\usepackage{floatrow}
\usepackage{slashed}

\usetikzlibrary{decorations.pathmorphing}

\newcommand\p{\partial}
\newcommand\rd{{\rm d}}

\newcommand{\der}{\partial}
\newcommand{\fr}{\frac}

\newcommand{\bs}{\boldsymbol}

\newcommand{\er}[1]{Eq.~\eqref{#1}}

\newcommand{\na}{\nabla}
\newcommand{\mtx}[1]{\left(\begin{matrix} #1 \end{matrix}\right)}

\begin{document}

\title{Self-similar inverse cascade from generalized symmetries}

\date{\today}

\author{Yuji~Hirono}
\email{yuji.hirono@gmail.com}
\affiliation{
    Institute of Systems and Information Engineering, University of Tsukuba, Tsukuba, Ibaraki 305-8573, Japan
}
\affiliation{Department of Physics, Osaka University, Toyonaka, Osaka 560-0043, Japan}

\author{Kohei Kamada}
\email{kohei.kamada@ucas.ac.cn}
\affiliation{School of Fundamental Physics and Mathematical Sciences, Hangzhou Institute for Advanced Study, University of Chinese Academy of Sciences (HIAS-UCAS), 310024 Hangzhou, China}
\affiliation{International Centre for Theoretical Physics Asia-Pacific (ICTP-AP), Hangzhou/Beijing, China}
\affiliation{Research Center for the Early Universe, The University of Tokyo, Bunkyo-ku, Tokyo 113-0033, Japan}

\author{Naoki Yamamoto}
\email{nyama@rk.phys.keio.ac.jp}
\affiliation{Department of Physics, Keio University, Yokohama 223-8522, Japan}
\author{Ryo Yokokura}
\email{ryokokur@keio.jp}
\affiliation{Department of Physics \& Research and Education Center for Natural Sciences, Keio University, Yokohama 223-8521, Japan}

\begin{abstract}
We investigate the role of generalized symmetries in driving non-equilibrium and non-linear phenomena, specifically focusing on turbulent systems. 
While conventional turbulence studies have revealed inverse cascades driven by conserved quantities integrated over the entire space, such as helicity in three spatial dimensions, the influence of higher-form symmetries, whose conserved charges are defined by integration over subspaces, remains largely unexplored.
We demonstrate a novel mechanism where higher-form symmetries naturally induce a self-similar inverse cascade. Taking axion electrodynamics with non-linear topological interaction as a paradigmatic example, we show that the conserved charge associated with its 1-form symmetry drives the system toward large-scale coherent structures through a universal scaling behavior characterized by analytically determined scaling exponents.
Our findings suggest that higher-form symmetries can provide a fundamental organizing principle for understanding non-equilibrium phenomena and the emergence of coherent structures in turbulent systems.
\end{abstract}

\maketitle

\sect{Introduction}%
Generalized symmetries~\cite{Gaiotto:2014kfa} have emerged as a powerful framework for understanding the fundamental structure of physical theories, extending beyond the conventional point-particle symmetries to include higher-form symmetries acting on extended objects such as loops or surfaces. These symmetries have provided crucial insights into quantum field theories including gauge theories~\cite{Gomes:2023ahz,Schafer-Nameki:2023jdn,Brennan:2023mmt,Bhardwaj:2023kri} and condensed matter systems~\cite{McGreevy:2022oyu}.
Their spontaneous breaking patterns and associated dynamics have been particularly important in understanding the organizing principles of various physical systems~\cite{Gaiotto:2014kfa,Lake:2018dqm,Wen:2018zux,Cordova:2018cvg,Hidaka:2020ucc,Qi:2020jrf,Hirono:2022dci}.

Despite the considerable attention these symmetries have received in vacuum (or equilibrium) physics, their role in non-equilibrium phenomena remains relatively unexplored. 
Higher-form symmetries have provided valuable insights into the formulation of magnetohydrodynamics (MHD)~\cite{Iqbal:2024pee} and analysis for the linear response of MHD~\cite{Das:2023nwl},
but their influence on strongly non-linear phenomena and turbulent behavior remains to be fully explored. 

In this Letter, we report a novel connection between higher-form symmetries and turbulent behavior,\footnote{We note that turbulence can arise in a wide range of systems governed by nonlinear field equations, not limited to hydrodynamic flows, with cascade phenomena serving as a hallmark of such turbulence. See, e.g., Ref.~\cite{Berges:2008mr} for studies of turbulence in non-Abelian gauge theories. The turbulence driven by the nonlinear topological interaction investigated in this work falls within this broader class.} specifically focusing on the emergence of inverse cascade phenomena~\cite{kraichnan1971inertial,fjortoft1953changes,tabeling2002two,alexakis2018cascades}.
While inverse cascades have been well established in systems with conserved charges defined by integration over the entire space domain, such as energy and enstrophy in two-dimensional turbulence~\cite{kraichnan1971inertial,fjortoft1953changes} or helicity in three-dimensional flows~\cite{moffatt1969degree,biferale2012inverse}, we demonstrate that higher-form symmetries, whose conserved charges are defined by integration over subspaces, can naturally induce a self-similar inverse cascade, where energy or conserved quantities flow hierarchically from small to large scales.
We exemplify this mechanism for axion electrodynamics with the non-linear topological interaction, which possesses a 1-form symmetry and associated conservation law~\cite{Sogabe:2019gif,Hidaka:2020iaz,Hidaka:2020izy,Choi:2022fgx,Yokokura:2022alv}. Our argument can also be applied to other systems with higher-form symmetries considered in, e.g., Ref.~\cite{Yamamoto:2023uzq}. 
This suggests a new paradigm for understanding the emergence of coherent structures in turbulent systems from fundamental properties of higher-form symmetries, with potential implications across a broad range of physical scenarios.

\sect{Higher-form symmetry and instability in axion electrodynamics}%
As a demonstration of our mechanism, we consider an example of the axion electrodynamics with a massless axion and a gauge field in $(3+1)$-dimensions. 
The action is given by
\begin{equation}
 S
= 
\int 
\rd^4 x 
\left( 
- \frac{v^2}2 \p_\mu  \phi \p^\mu \phi 
- \frac{1}{4e^2} f_{\mu \nu } f^{\mu\nu} 
- 
\frac{C}{4}
\phi 
f_{\mu \nu} \tilde{f}^{\mu \nu }
\right) ,
\end{equation}
where $\phi$ is a $2\pi $ periodic pseudo-scalar field identified as an axion, $f_{\mu \nu } = \der_\mu  a_\nu - \der_\nu  a_\mu$ is the field strength of a $U(1)$ gauge field $a_\mu $, $v$ is a parameter with mass dimension 1, $e$ is a coupling constant, and $C=1/(4\pi^2)$.
The symbol $\tilde{f}^{\mu \nu } = \fr{1}{2} 
\epsilon^{\mu \nu\rho \sigma} f_{\rho\sigma}$
with the totally anti-symmetric tensor $\epsilon^{\mu\nu\rho\sigma}$ denotes the dual of the field strength.
In the following, 
we consider the dynamics of 
the system as a classical field theory.
The equations of motion of this system read
\begin{equation}
    \begin{split}
        v^2 \der_\mu \der^\mu \phi 
        - \fr{C}{4} f_{\mu \nu} \tilde{f}^{\mu \nu} &= 0 ,
        \\
       \fr{1}{e^2} \der_\mu f^{\mu \nu}
        + C (\der_\mu \phi ) \tilde{f}^{\mu \nu} &= 0. 
    \end{split}
\end{equation}

The system has been shown to exhibit instability in the presence of background electric fields~\cite{Bergman:2011rf, Ooguri:2011aa, Yamamoto:2022vrh, Yamamoto:2023uzq}. 
Under a uniform nonzero electric field
$f_{03} (0, \bs{x})= - \bar{E}_3$,
the dispersion relation for modes with momentum $\bs{k} = (k_1, 0, 0)$ takes the form
\begin{equation}
    \omega^2 = k_1^2 \pm \fr{C e \bar{E}_3}{v}|k_1|.
\label{dispersion}
\end{equation}
This reveals an unstable mode in the infrared region $0 < |k_1| < C e \bar{E}_3/v$.

Through the conservation laws, we can demonstrate that this instability necessarily reduces the electric field.
The conserved charges associated with the 0-form symmetry $\phi \rightarrow \phi + \lambda$ and 1-form symmetry $a_{\mu} \rightarrow a_{\mu} + \lambda_{\mu}$ are
\begin{equation}
\begin{split}
Q_\phi ({\cal V})
&=K_\phi ({\cal V}) + T_\phi ({\cal V}), \\
Q_a ({\cal S})
&=K_a ({\cal S}) + T_a ({\cal S}), 
\end{split}
\end{equation}
respectively, where we have defined 
\begin{align}
 K_\phi ({\cal V})  = 
\int_{\cal V} \rd \tilde{V}_\mu v^2 \der^\mu \phi ,
\quad
&
T_\phi ({\cal V})   = 
- 
\int_{\cal V} \rd \tilde{V}_\mu
\fr{C}{4} a_\nu  \tilde{f}^{\mu \nu},
\\
 K_a ({\cal S}) = \fr{1}{2}
\int_{\cal S} \rd S^{\mu\nu}
\fr{1}{e^2} 
\tilde{f}_{\mu \nu},
\quad
&
T_a  ({\cal S})   = 
 - \fr{1}{2} \int_{\cal S} \rd S^{\mu\nu}
C \phi f_{\mu \nu},
\end{align}
with ${\cal V}$ and ${\cal S}$ being 3- and 2-dimensional closed subspaces.
We have decomposed the conserved charges into those from the kinetic terms, $K_\phi ({\cal V})$ and $ K_a ({\cal S})$, and those from the topological terms, $T_\phi ({\cal V})$ and $T_a ({\cal S})$.
In the following, we assume that monopoles and axionic strings are sufficiently heavy so that they do not appear dynamically, ensuring that the associated conserved charges are well defined.%
\footnote{It should be noted that, at the quantum level,
$T_\phi ({\cal V})$ and $T_a ({\cal S})$ are not invariant under large gauge transformations. They can be made invariant under those transformations by rewriting them in terms of the topological quantum field theories (TQFT), and the resulting symmetries become non-invertible~\cite{Choi:2022jqy, Cordova:2022ieu,Choi:2022fgx,Yokokura:2022alv, Yamamoto:2023uzq}. However, as we will not consider the backgrounds with 't Hooft lines or axionic vortices, this reformulation will not matter in the following discussion, and we will write these topological quantities without introducing TQFT.}
\footnote{We also have higher-form symmetries associated with the Bianchi identities $\partial_\mu \tilde{f}^{\mu\nu} =0$
and $\epsilon^{\mu \nu \rho \sigma}\der_\rho \der_\sigma  \phi =0$. Since these conservation laws do not involve the interaction terms, they do not contribute to the inverse cascade.
Meanwhile, the presence of these symmetries implies the absence of dynamical magnetic monopoles and axionic strings.}

While $Q_\phi$ is the total helicity consisting of the chiral charge $K_\phi$ carried by axions and the usual magnetic helicity $T_\phi$,
$Q_a$ is a conserved charge associated with the 1-form symmetry consisting of the electric flux $K_a$ and the topological charge $T_a$. 
The charge $T_a$ has a geometric meaning of the linking number, similar to the usual magnetic helicity~\cite{Yamamoto:2023uzq}.
By taking ${\cal S} $ as an $x^1 x^2$-plane with $x^3 =0$ at the time $x^0$ denoted by $S_{12} (x^0) $, we have 
\begin{equation}
\label{Q_a}
 Q_a (S_{12}  (x^0) )
= 
- \int_{S_{12} (x^0) } \rd S^{12}
\left( \fr{1}{e^2} f_{03} + C \phi f_{12} 
\right).
\end{equation}
At initial time $x^0 =0$, the surface integral is dominated by $K_a (S_{12} (0))$ with the initial electric field
$f_{03} = - \bar{E}_3$.
Meanwhile, the tachyonic growth of $\phi$ and $f_{12}$ generates the topological charge $ T_a(S_{12} (x^0))$. 
Consequently, the conservation law dictates that $K_a(S_{12} (x^0))$ and electric field $f_{03}$ decrease.

To understand the physical nature of this mechanism, it is instructive to compare it with the chiral plasma instability (CPI) that occurs in the presence of fermion chirality imbalance~\cite{Joyce:1997uy, Akamatsu:2013pjd}. In the CPI, the generation of magnetic helicity leads to a reduction of the initial chirality imbalance due to total helicity conservation. In our case, the role of helicity conservation is replaced by the conservation law associated with the 1-form symmetry, suggesting this can be viewed as a generalized chiral instability~\cite{Yamamoto:2023uzq}.

\sect{Dissipation}%
In realistic situations in a medium, we must account for dissipative effects. 
As a concrete setup, we consider a system coupled to gapped charged matter, where the Ohmic current is absent. 
However, the axion field can still experience dissipation through its interactions with environmental degrees of freedom. 
Following the bottom-up approach to low-energy effective theory, we include the leading-order dissipative term allowed by symmetry in the axion sector.\footnote{
While a full treatment within the Schwinger-Keldysh effective field theory framework~\cite{Crossley:2015evo, Liu:2018kfw} would clarify the associated Kubo-Martin-Schwinger constraints and confirm the absence of other symmetry-allowed terms at the same order, such an analysis is beyond the scope of the present work. We therefore treat this term as a phenomenological assumption consistent with symmetry. 
} The corresponding low-energy equation of motion for the axion is given by
\begin{align}
v^2 (- \der_0^2
+ \na^2) \phi 
- \gamma \der_0 \phi 
        - \fr{C}{4} f_{\mu \nu} \tilde{f}^{\mu \nu} 
        = 0,
\label{EOM_aI}
\end{align}
where $\gamma$ is the diffusion coefficient.
Importantly, the presence of $\gamma$ preserves both the unstable mode and the conservation law~\eqref{Q_a}. 
This preservation of the conservation law in the presence of dissipation sets the stage for a novel cascade mechanism.

\sect{Mechanism of inverse cascade}%
The interplay between dissipation and charge conservation provides a fundamental mechanism for the inverse cascade. 
The dissipation term $\gamma \p_0 \phi$ reduces the total energy of the system over time. 
However, the conserved charge $Q_a = K_a + T_a$ must remain constant due to the 1-form symmetry. 
Initially, $Q_a$ is dominated by the kinetic component $K_a$ from the electric field. 
As the instability develops, this charge is transferred to the topological component $T_a$, which eventually dominates the late-time dynamics. 
Since the charge density at wavenumber $\sim k$ contributes energy $\sim k^2$ through spatial derivatives, maintaining constant charge while minimizing energy necessarily drives the system toward smaller $k$ modes. This mechanism continuously transfers the topological charge to larger scales.

Our numerical simulation confirms this inverse cascade mechanism. Additionally, we observe the emergence of self-similar scaling behavior, where energy and topological charge are hierarchically transferred to increasingly larger scales. As we demonstrate through scaling analysis below, this leads to universal scaling exponents $\alpha = 1, \beta = 1/2$ characterizing the late-time evolution of the system.

\begin{figure*}[ht!]
    \centering
     \subfigure[\ Spectral functions for energy $f_\phi(t,k)+f_{a_2}(t,k)$.]{
      \includegraphics[width=0.39\columnwidth]{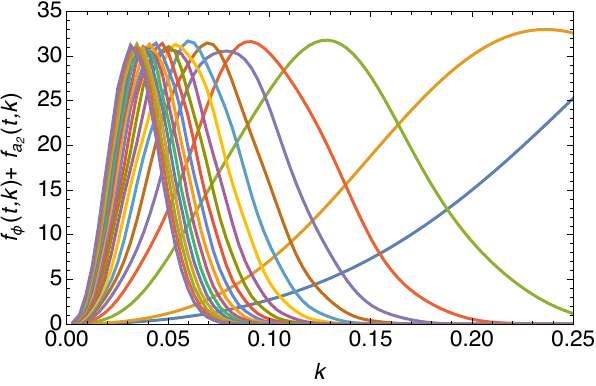} 
  \label{fig:spec} 
    }
    \subfigure[\ Spectral functions for topological charge $g(t,k)$.]{
\includegraphics[width=0.57\columnwidth]{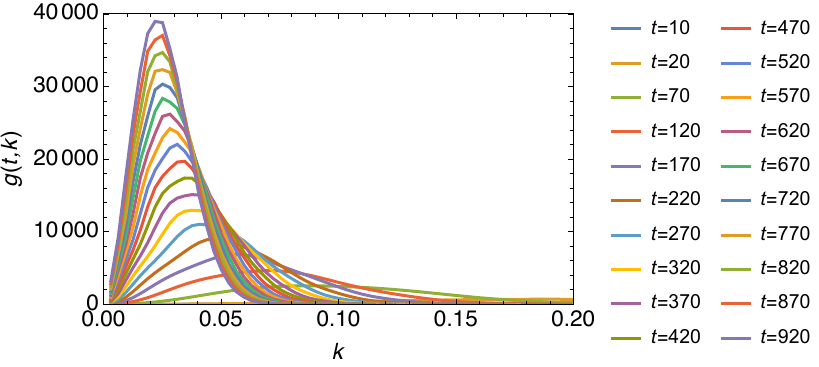} 
    }    
  \vspace{-2mm}
    \caption{\ Spectral functions for energy and topological charge at different times. 
    Both spectral functions exhibit the inverse cascade.} 
    \label{fig:spec-functions}
\end{figure*}

\sect{Set up for numerical simulations}%
To demonstrate the emergence of self-similar inverse cascade, we numerically analyze the long-time behavior of the system. While the full $(3+1) $-dimensional dynamics involves complex interplay between all field components, the essential mechanism of generalized symmetry-induced cascade can be captured by studying field variations along a single direction.%
\footnote{A similar approximation is employed in the analysis of the CPI in the presence of magnetic fields~\cite{Hirono:2015rla}.}
Specifically, we assume the system is homogeneous along the $x^2$- and $x^3$-directions, and the field $a_3$ is spatially homogeneous, 
$a_3 (x^\mu ) = a_3 (x^0)$.\footnote{While this dimensional reduction simplifies the numerical analysis, we expect that the essential features of the inverse cascade should persist in the full (3+1)-dimensional case. This expectation is based on the general argument given earlier, where the interplay between dissipation and the conservation of the 1-form charge drives the transfer of the conserved quantity to larger spatial scales. This mechanism does not rely on the dimensional reduction and should apply in the full theory as well.}
Under these conditions and in the temporal gauge $a_0 =0$, the equations of motion reduce to
\begin{align}
v^2 ( - \der_0^2 \phi + \der_1^2 \phi )
- \gamma \der_0 \phi
       &
= - C \der_0 a_3 \der_1 a_2
,
\label{EOM_phi}
\\
\der_0 f^{01} & = 0,
\label{EOM_a1}
\\
\fr{1}{e^2}(- \der_0^2 a_2 + \der_1^2 a_2)
& = 
C
\der_1  \phi  \der_0 a_3
,
\label{EOM_a2}
\\
- 
\fr{1}{e^2}  \der_0^2 a_3 
 & = 
C \left(
\der_0  \phi  \der_1 a_2 
- 
\der_1  \phi \der_0 a_2
\right).
\label{EOM_a3}
\end{align}

We solve the system in a box, $x^1 \in [-L, L]$, with the periodic boundary condition.
First, we focus on the conservation law obtained by
\er{EOM_a3}. 
The conserved charge integrated along the $x^1$-direction can be written as 
\begin{equation}
\label{N}
N =
\int_{-L}^L \rd x^1 
\left( 
\fr{1}{e^2}\der_0 a_3  + C \phi \der_1 a_2
\right)
= 
\fr{2L v}{ C e^3}
A(x^0)
+ N_{\rm top}\,,
\end{equation}
where 
\begin{equation}
A(x^0) \coloneqq \frac{C e}{v} \der_0 a_3 (x^0)
\end{equation}
is proportional to 
the spatial average of the electric field, 
\begin{equation}
N_{\rm top}    
\coloneqq C \int_{-L}^L \rd x^1 \phi \der_1 a_2
\end{equation}
is the topological charge, and $Q_a (S_{12}(x^0)) = \int \rd x^2 N$.
The quantity $A(x^0)$ determines the scale below which the modes exhibit instability as in \er{dispersion}.

\begin{figure*}[ht!]
    \centering
     \subfigure[\ Rescaled spectral functions for energy $f_\phi(t,k)+f_{a_2}(t,k)$.]{
      \includegraphics[width=0.4\columnwidth]{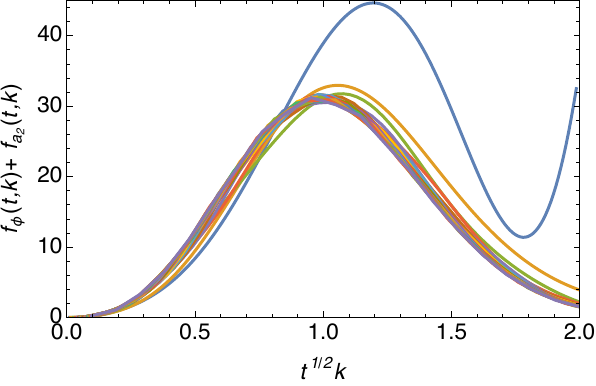} 
    }
    \subfigure[\ Rescaled spectral functions for topological charge $g(t,k)$.]{
\includegraphics[width=0.56\columnwidth]{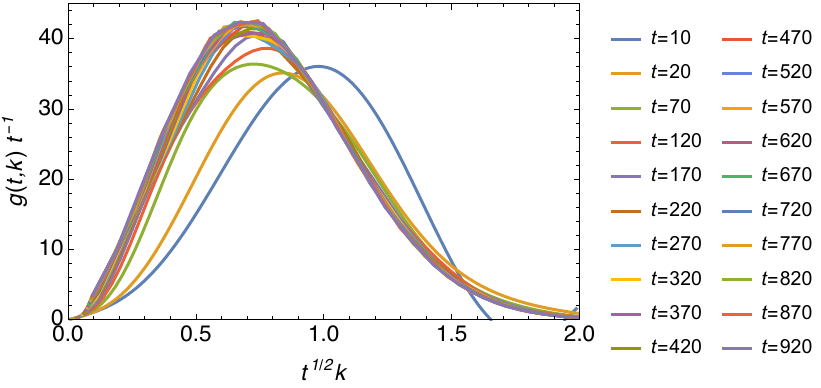} 
  \label{fig:rescaled-spec} 
    }    
  \vspace{-2mm}
    \caption{\ Rescaled spectral functions for energy and topological charge at different times according to the scaling laws~\eqref{eq:exponents}--\eqref{eq:gscaling}. 
    Both rescaled spectral functions converge at later times, indicating self-similarity.} 
\end{figure*}

To simplify our notation, we introduce the coordinates $(t, x) \coloneqq (x^0 , x^1) $, dimensionless fields $(b, c) \coloneqq (e^2 \phi, {e}a_2/{v})$, and the rescaled parameter $\sigma \coloneqq {\gamma}/{v^2} $.
The equations of motion for $b$ and $c$ then take the form
\begin{equation}
\der_t^2 \mtx{b \\ c}
+ 
\sigma \der_t \mtx{b \\ 0}
=
\mtx{\der_x^2 & A \der_x \\ 
-A \der_x & \der_x^2
}
\mtx{b \\ c}
.
\label{eq:eom-bc-mat}
\end{equation}
To solve Eq.~\eqref{eq:eom-bc-mat} in momentum space, we expand $b(t,x)$ and $c(t,x)$ as 
\begin{align}
b(t,x) \! &= \! \frac{b_{\rm e}(t,k_0)}2 \! + \! \sum_{n=1}^\infty 
\left(
b_{\rm e}(t,k_n) \cos k_n x \! + \! b_{\rm o} (t,k_n)\sin k_n x
\right), 
\nonumber
\\
c(t,x) \! &= \! \frac{c_{\rm e}(t,k_0)}2 \! + \! \sum_{n=1}^\infty
\left(
c_{\rm e} (t,k_n)\cos k_n x \! + \! c_{\rm o}(t,k_n) \sin k_n x
\right),
\nonumber
\end{align}
where discrete momenta are given by $k_n = {\pi n}/{L}$.

\sect{Numerical results}%
We numerically solve Eq.~\eqref{eq:eom-bc-mat} in momentum space for $\{b_{\rm e,o}(t,k_n), c_{\rm e,o}(t,k_n) \}_{n=1,\ldots,N_{k}}$, where $N_k$ sets the UV cutoff.
Our analysis uses the parameters
$L = 1000 $, 
$N_k= 200 $, $\sigma = 1.0$,
$e = 1.0$, and $v= 1$, 
with initial conditions
$N  = 24 \pi^2 L $ and 
$(b_{\rm e,o}(0,k_n), c_{\rm e,o}(0,k_n)) = (0,2,0, 0)$
for $n=1,\ldots,N_{k}$.
The initial value $A(0)$ is given by
$A (0) = {C e^3} N/({2 L})  \simeq 3.0$.

We now analyze the time evolution of physical quantities. The energy of the $\phi$ and $a_2$ fields and the topological charge can be expressed using the Fourier coefficients $\{b_{\rm e,o}(t,k_n), c_{\rm e,o}(t,k_n)\}_{n=1,\ldots}$ as
\begin{align}
\begin{split}
E_X 
= \frac{v^2 L}{2e^2} \sum_{n=1}^{\infty} f_X(t,k_n), \quad  X=\phi,a_2, 
\end{split}\\
\begin{split}    
N_{\rm top} = 
\frac{C L v}{e^3}  
\sum_{n=1}^\infty k_n g(t,k_n),
\end{split}
\end{align}
where we introduced spectral functions by
\begin{align}
f_\phi (t, k_n) &\coloneqq  \sum_{i={\rm e,o}} 
\left[(\partial_t b_i(t,k_n))^2 + k_n^2 (b_i(t,k_n))^2 \right], \\
f_{a_2} (t, k_n)&\coloneqq \sum_{i={\rm e,o}} 
\left[(\partial_t c_i(t,k_n))^2 + k_n^2 (c_i(t,k_n))^2 \right], \\
g(t, k_n) &\coloneqq  b_{\rm e} (t,k_n)c_{\rm o}(t,k_n) - b_{\rm o}(t,k_n) c_{\rm e}(t,k_n) .
\end{align}
In Fig.~\ref{fig:spec-functions}, we plot the spectral functions $f_\phi(t,k)+f_{a_2}(t,k)$ and $g(t,k)$ of the energy and topological charge, respectively, at different times. 
The results show that the $a_2$ and $\phi$ fields, initially amplified by the generalized chiral instability, develop inverse cascades at later times.

\sect{Self-similar scaling induced by higher-form symmetries}%
The late-time dynamics of the system exhibits self-similar behavior, which we characterize through scaling analysis.
We propose that the Fourier components take the scaling form
\begin{equation}
b_i(t,k) \sim t^{\alpha/2} \tilde{b}_i(t^\beta k), \quad c_i(t,k) \sim t^{\alpha/2} \tilde{c}_i(t^\beta k),
\label{eq:scaling-ansatz}
\end{equation}
for $i={\rm e,o}$. The exponents $\alpha$ and $\beta$ can be determined from the equations of motion and conservation laws, as we show below.
The scaling ansatz~\eqref{eq:scaling-ansatz} implies that the spectral function $g(t,k)$ obeys 
\begin{equation}
g(t,k) \sim t^\alpha \tilde{g}(t^\beta k).
\end{equation}

To find $\beta$, we examine the structure of Eq.~\eqref{eq:eom-bc-mat}. 
In the late-time regime, the first-order time derivative term becomes dominant in Eq.~\eqref{eq:eom-bc-mat}.
The $(\p_x)^2$ term generates time dependence of the form $t k^2 = (t^{1/2}k)^2$ upon integration, requiring $\beta = 1/2$. For consistency, the $A \p_x$ term, which contributes $k \int^t A(t') \rd t'$ must also scale as $t^{1/2} k$. This consistency condition determines the scaling of $A(t)$:
\begin{equation}
\label{A_scaling}
A(t) \sim t^{-1/2}.
\end{equation}
The decay of $A(t)$ indicates that the conserved charge $N$ becomes dominated by its topological component $N_{\rm top}$ at late times. In the scaling regime, this topological charge takes the form
\begin{equation}
\int \rd k\, k g(t,k) 
= t^{\alpha - 2\beta} \int \rd y\, \tilde{g}(y),
\end{equation}
where $y:=t^{\beta}k$.
The constancy of this quantity requires $\alpha = 2\beta$, yielding our final scaling exponents:
\begin{equation}
\alpha = 1, \quad \beta = \frac{1}2\,.
\label{eq:exponents}
\end{equation}
These scaling exponents lead to the following relations for the spectral functions:
\begin{equation}\label{eq:fscaling}
f_\phi(t,k) 
\sim \sum_{i={\rm e}, {\rm o}}
(k b_i(t,k))^2 \sim y^2 \tilde{f}_\phi(y) 
\end{equation}
and analogous scaling for $f_{a_2}(t,k)$, and
\begin{equation}\label{eq:gscaling}
g(t,k) \sim t \, \tilde{g}(y).
\end{equation}
Note that the kinetic energies in $f_\phi(t,k)$ and $f_{a_2}(t,k)$ scale as $\propto t^{-1}$ and become subdominant. 

\begin{figure}[t]
\begin{center}
  \includegraphics
  [scale = 0.7]
  {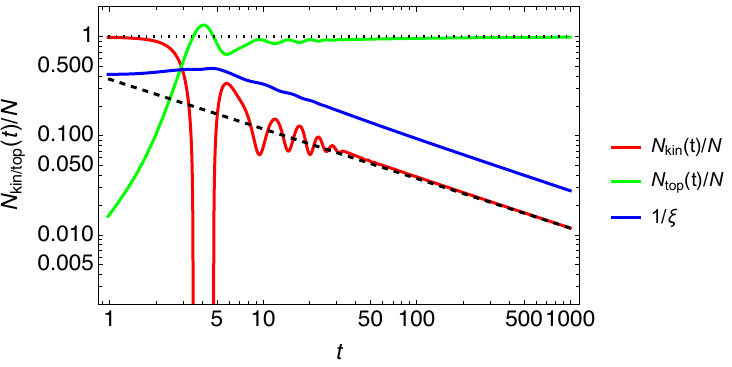} 
  \end{center}
  \caption{
  \label{fig:top}  Time evolution of $N_{\rm kin}$ (red), $N_{\rm top}$ (green), and $\xi^{-1}$ (blue). Black dashed line illustrates $t^{-1/2}$ asymptotic behavior of $N_{\rm kin}$ and $\xi^{-1}$.
  }
\end{figure}

These theoretical predictions are confirmed by our numerical simulations. Figure~\ref{fig:rescaled-spec} shows the spectral function rescaled according to Eq.~\eqref{eq:exponents}, demonstrating clear convergence to a universal curve at late times. 
Figure~\ref{fig:top} shows the time evolution of the charges stored in $N_{\rm kin}$ and $N_{\rm top}$
as well as the inverse of the coherence length $\xi$ defined by 
\begin{equation}
\xi \coloneqq \frac{\sum_n  k_n^{-1} (f_\phi(t,k_n) +f_{a_2}(t,k_n))  }{\sum_n (f_\phi(t,k_n) +f_{a_2}(t,k_n)) }. 
\end{equation}
Both $N_{\rm kin}$ (or equivalently, $A(t)$) and $\xi^{-1}$ show asymptotic behavior $t^{-1/2}$, consistent with the predicted scaling~\eqref{A_scaling}, further validating our scaling analysis.

\sect{Summary and discussion}%
In this Letter, we have demonstrated that higher-form symmetries can induce self-similar inverse cascade in a turbulent system. Using axion electrodynamics as a concrete example, we showed that starting from an initial electric field configuration that triggers the generalized chiral instability, the system exhibits characteristic scaling behavior at late times. 
In particular, the spectral function $g(t,k)$ grows according to the scaling law $g(lt,k) \sim l g(t, l^{1/2}k)$, while the instability parameter $A(t)$ decays as $t^{-1/2}$. 
The inverse cascade, in which spectral weight is transferred from short to long wavelengths, is ensured by the conservation law associated with the 1-form symmetry together with energy dissipation. 
To determine the scaling exponents, we imposed a self-similar ansatz and showed that the equations of motion admit such a solution. While the existence of a self-similar solution itself is non-trivial, it remains unclear a priori whether this solution is dynamically realized or acts as an attractor of the evolution. Our numerical simulations confirm that the system indeed approaches this self-similar scaling regime at late times.

Understanding the ultimate state of the system after the inverse-cascade scaling remains an open problem for future work.
In the case of two-dimensional fluid turbulence in a finite system, which also exhibits an inverse cascade, it was initially conjectured that energy accumulates in low-energy modes, analogous to Bose-Einstein condensation~\cite{kraichnan1967}. This conjecture has since been examined in both numerical simulations and experiments~\cite{boffetta2012two}. A similar phenomenon may also be expected for the inverse-cascading turbulence studied here under finite-size conditions, although further investigation will be required.

The present results indicate the possible existence of a new universality class protected by higher-form symmetries. Interestingly, the scaling exponents we obtained coincide with those found in turbulence after the CPI~\cite{Tashiro:2012mf,Hirono:2015rla,Yamamoto:2016xtu,Buividovich:2015jfa,Mace:2019cqo}. 
This coincidence may suggest that universality classes protected by 0-form and 1-form symmetries are connected within a broader framework. Whether such a universality indeed exists, and how these connections are realized, remain open questions for future studies.

The phenomena studied here may also be realized in concrete physical systems. One promising class of candidates lies in condensed matter experiments, such as magnetic materials where magnetic fluctuations can act as emergent axion fields~\cite{Ooguri:2011aa}. Another natural arena is cosmology. In particular, similar dynamics may manifest in gauge field amplification during inflation~\cite{Turner:1987bw,Ratra:1991bn} in the presence of spectator axion-like fields.

Our analysis can be extended to the full $(3+1)$-dimensional axion electrodynamics to uncover additional features of the cascade process. Furthermore, exploring connections between our findings and other non-equilibrium systems with higher-form symmetries may reveal new universal aspects of turbulent behavior.

Although we have not included monopoles and axionic strings in our analysis, it would be non-trivial how their existence affects the dynamics, since they would violate the conservation laws of $Q_\phi$, $Q_a$, and magnetic symmetries.
Similar considerations also apply to the conventional CPI.
We leave this problem for future study.

\sect{Acknowledgments}%
This work was supported in part by JSPS KAKENHI Grant Numbers 
JP22H05111 (Y.\,H.), JP22H05118 (Y.\,H.), JP24K23186 (Y.\,H.),
JP23K17687 (K.\,K.), 
JP22H01216 (N.\,Y.), JP24K00631 (N.\,Y.), 
JP21K13928 (R.\,Y.), JP25K17394 (R.\,Y.),
JST, PRESTO Grant Number JPMJPR24K8 (Y.\,H.), 
and the National Natural Science Foundation of China (NSFC) under Grant Number 12547104 (K.\,K.).

\bibliography{inverse-cascade}

\begin{thebibliography}{45}%
\makeatletter
\providecommand \@ifxundefined [1]{%
 \@ifx{#1\undefined}
}%
\providecommand \@ifnum [1]{%
 \ifnum #1\expandafter \@firstoftwo
 \else \expandafter \@secondoftwo
 \fi
}%
\providecommand \@ifx [1]{%
 \ifx #1\expandafter \@firstoftwo
 \else \expandafter \@secondoftwo
 \fi
}%
\providecommand \natexlab [1]{#1}%
\providecommand \enquote  [1]{``#1''}%
\providecommand \bibnamefont  [1]{#1}%
\providecommand \bibfnamefont [1]{#1}%
\providecommand \citenamefont [1]{#1}%
\providecommand \href@noop [0]{\@secondoftwo}%
\providecommand \href [0]{\begingroup \@sanitize@url \@href}%
\providecommand \@href[1]{\@@startlink{#1}\@@href}%
\providecommand \@@href[1]{\endgroup#1\@@endlink}%
\providecommand \@sanitize@url [0]{\catcode `\\12\catcode `\$12\catcode
  `\&12\catcode `\#12\catcode `\^12\catcode `\_12\catcode `\%12\relax}%
\providecommand \@@startlink[1]{}%
\providecommand \@@endlink[0]{}%
\providecommand \url  [0]{\begingroup\@sanitize@url \@url }%
\providecommand \@url [1]{\endgroup\@href {#1}{\urlprefix }}%
\providecommand \urlprefix  [0]{URL }%
\providecommand \Eprint [0]{\href }%
\providecommand \doibase [0]{http://dx.doi.org/}%
\providecommand \selectlanguage [0]{\@gobble}%
\providecommand \bibinfo  [0]{\@secondoftwo}%
\providecommand \bibfield  [0]{\@secondoftwo}%
\providecommand \translation [1]{[#1]}%
\providecommand \BibitemOpen [0]{}%
\providecommand \bibitemStop [0]{}%
\providecommand \bibitemNoStop [0]{.\EOS\space}%
\providecommand \EOS [0]{\spacefactor3000\relax}%
\providecommand \BibitemShut  [1]{\csname bibitem#1\endcsname}%
\let\auto@bib@innerbib\@empty
\bibitem [{\citenamefont {Gaiotto}\ \emph {et~al.}(2015)\citenamefont
  {Gaiotto}, \citenamefont {Kapustin}, \citenamefont {Seiberg},\ and\
  \citenamefont {Willett}}]{Gaiotto:2014kfa}%
  \BibitemOpen
  \bibfield  {author} {\bibinfo {author} {\bibfnamefont {D.}~\bibnamefont
  {Gaiotto}}, \bibinfo {author} {\bibfnamefont {A.}~\bibnamefont {Kapustin}},
  \bibinfo {author} {\bibfnamefont {N.}~\bibnamefont {Seiberg}}, \ and\
  \bibinfo {author} {\bibfnamefont {B.}~\bibnamefont {Willett}},\ }\href
  {\doibase 10.1007/JHEP02(2015)172} {\bibfield  {journal} {\bibinfo  {journal}
  {JHEP}\ }\textbf {\bibinfo {volume} {02}},\ \bibinfo {pages} {172} (\bibinfo
  {year} {2015})},\ \Eprint {http://arxiv.org/abs/1412.5148} {arXiv:1412.5148
  [hep-th]} \BibitemShut {NoStop}%
\bibitem [{\citenamefont {Gomes}(2023)}]{Gomes:2023ahz}%
  \BibitemOpen
  \bibfield  {author} {\bibinfo {author} {\bibfnamefont {P.~R.~S.}\
  \bibnamefont {Gomes}},\ }\href {\doibase 10.21468/SciPostPhysLectNotes.74}
  {\bibfield  {journal} {\bibinfo  {journal} {SciPost Phys. Lect. Notes}\
  }\textbf {\bibinfo {volume} {74}},\ \bibinfo {pages} {1} (\bibinfo {year}
  {2023})},\ \Eprint {http://arxiv.org/abs/2303.01817} {arXiv:2303.01817
  [hep-th]} \BibitemShut {NoStop}%
\bibitem [{\citenamefont {Schafer-Nameki}(2024)}]{Schafer-Nameki:2023jdn}%
  \BibitemOpen
  \bibfield  {author} {\bibinfo {author} {\bibfnamefont {S.}~\bibnamefont
  {Schafer-Nameki}},\ }\href {\doibase 10.1016/j.physrep.2024.01.007}
  {\bibfield  {journal} {\bibinfo  {journal} {Phys. Rept.}\ }\textbf {\bibinfo
  {volume} {1063}},\ \bibinfo {pages} {1} (\bibinfo {year} {2024})},\ \Eprint
  {http://arxiv.org/abs/2305.18296} {arXiv:2305.18296 [hep-th]} \BibitemShut
  {NoStop}%
\bibitem [{\citenamefont {Brennan}\ and\ \citenamefont
  {Hong}(2023)}]{Brennan:2023mmt}%
  \BibitemOpen
  \bibfield  {author} {\bibinfo {author} {\bibfnamefont {T.~D.}\ \bibnamefont
  {Brennan}}\ and\ \bibinfo {author} {\bibfnamefont {S.}~\bibnamefont {Hong}},\
  }\href@noop {} {\  (\bibinfo {year} {2023})},\ \Eprint
  {http://arxiv.org/abs/2306.00912} {arXiv:2306.00912 [hep-ph]} \BibitemShut
  {NoStop}%
\bibitem [{\citenamefont {Bhardwaj}\ \emph {et~al.}(2024)\citenamefont
  {Bhardwaj}, \citenamefont {Bottini}, \citenamefont {Fraser-Taliente},
  \citenamefont {Gladden}, \citenamefont {Gould}, \citenamefont {Platschorre},\
  and\ \citenamefont {Tillim}}]{Bhardwaj:2023kri}%
  \BibitemOpen
  \bibfield  {author} {\bibinfo {author} {\bibfnamefont {L.}~\bibnamefont
  {Bhardwaj}}, \bibinfo {author} {\bibfnamefont {L.~E.}\ \bibnamefont
  {Bottini}}, \bibinfo {author} {\bibfnamefont {L.}~\bibnamefont
  {Fraser-Taliente}}, \bibinfo {author} {\bibfnamefont {L.}~\bibnamefont
  {Gladden}}, \bibinfo {author} {\bibfnamefont {D.~S.~W.}\ \bibnamefont
  {Gould}}, \bibinfo {author} {\bibfnamefont {A.}~\bibnamefont {Platschorre}},
  \ and\ \bibinfo {author} {\bibfnamefont {H.}~\bibnamefont {Tillim}},\ }\href
  {\doibase 10.1016/j.physrep.2023.11.002} {\bibfield  {journal} {\bibinfo
  {journal} {Phys. Rept.}\ }\textbf {\bibinfo {volume} {1051}},\ \bibinfo
  {pages} {1} (\bibinfo {year} {2024})},\ \Eprint
  {http://arxiv.org/abs/2307.07547} {arXiv:2307.07547 [hep-th]} \BibitemShut
  {NoStop}%
\bibitem [{\citenamefont {McGreevy}(2023)}]{McGreevy:2022oyu}%
  \BibitemOpen
  \bibfield  {author} {\bibinfo {author} {\bibfnamefont {J.}~\bibnamefont
  {McGreevy}},\ }\href {\doibase 10.1146/annurev-conmatphys-040721-021029}
  {\bibfield  {journal} {\bibinfo  {journal} {Ann. Rev. Condensed Matter
  Phys.}\ }\textbf {\bibinfo {volume} {14}},\ \bibinfo {pages} {57} (\bibinfo
  {year} {2023})},\ \Eprint {http://arxiv.org/abs/2204.03045} {arXiv:2204.03045
  [cond-mat.str-el]} \BibitemShut {NoStop}%
\bibitem [{\citenamefont {Lake}(2018)}]{Lake:2018dqm}%
  \BibitemOpen
  \bibfield  {author} {\bibinfo {author} {\bibfnamefont {E.}~\bibnamefont
  {Lake}},\ }\href@noop {} {\  (\bibinfo {year} {2018})},\ \Eprint
  {http://arxiv.org/abs/1802.07747} {arXiv:1802.07747 [hep-th]} \BibitemShut
  {NoStop}%
\bibitem [{\citenamefont {Wen}(2019)}]{Wen:2018zux}%
  \BibitemOpen
  \bibfield  {author} {\bibinfo {author} {\bibfnamefont {X.-G.}\ \bibnamefont
  {Wen}},\ }\href {\doibase 10.1103/PhysRevB.99.205139} {\bibfield  {journal}
  {\bibinfo  {journal} {Phys. Rev. B}\ }\textbf {\bibinfo {volume} {99}},\
  \bibinfo {pages} {205139} (\bibinfo {year} {2019})},\ \Eprint
  {http://arxiv.org/abs/1812.02517} {arXiv:1812.02517 [cond-mat.str-el]}
  \BibitemShut {NoStop}%
\bibitem [{\citenamefont {C\'ordova}\ \emph {et~al.}(2019)\citenamefont
  {C\'ordova}, \citenamefont {Dumitrescu},\ and\ \citenamefont
  {Intriligator}}]{Cordova:2018cvg}%
  \BibitemOpen
  \bibfield  {author} {\bibinfo {author} {\bibfnamefont {C.}~\bibnamefont
  {C\'ordova}}, \bibinfo {author} {\bibfnamefont {T.~T.}\ \bibnamefont
  {Dumitrescu}}, \ and\ \bibinfo {author} {\bibfnamefont {K.}~\bibnamefont
  {Intriligator}},\ }\href {\doibase 10.1007/JHEP02(2019)184} {\bibfield
  {journal} {\bibinfo  {journal} {JHEP}\ }\textbf {\bibinfo {volume} {02}},\
  \bibinfo {pages} {184} (\bibinfo {year} {2019})},\ \Eprint
  {http://arxiv.org/abs/1802.04790} {arXiv:1802.04790 [hep-th]} \BibitemShut
  {NoStop}%
\bibitem [{\citenamefont {Hidaka}\ \emph
  {et~al.}(2021{\natexlab{a}})\citenamefont {Hidaka}, \citenamefont {Hirono},\
  and\ \citenamefont {Yokokura}}]{Hidaka:2020ucc}%
  \BibitemOpen
  \bibfield  {author} {\bibinfo {author} {\bibfnamefont {Y.}~\bibnamefont
  {Hidaka}}, \bibinfo {author} {\bibfnamefont {Y.}~\bibnamefont {Hirono}}, \
  and\ \bibinfo {author} {\bibfnamefont {R.}~\bibnamefont {Yokokura}},\ }\href
  {\doibase 10.1103/PhysRevLett.126.071601} {\bibfield  {journal} {\bibinfo
  {journal} {Phys. Rev. Lett.}\ }\textbf {\bibinfo {volume} {126}},\ \bibinfo
  {pages} {071601} (\bibinfo {year} {2021}{\natexlab{a}})},\ \Eprint
  {http://arxiv.org/abs/2007.15901} {arXiv:2007.15901 [hep-th]} \BibitemShut
  {NoStop}%
\bibitem [{\citenamefont {Qi}\ \emph {et~al.}(2021)\citenamefont {Qi},
  \citenamefont {Radzihovsky},\ and\ \citenamefont {Hermele}}]{Qi:2020jrf}%
  \BibitemOpen
  \bibfield  {author} {\bibinfo {author} {\bibfnamefont {M.}~\bibnamefont
  {Qi}}, \bibinfo {author} {\bibfnamefont {L.}~\bibnamefont {Radzihovsky}}, \
  and\ \bibinfo {author} {\bibfnamefont {M.}~\bibnamefont {Hermele}},\ }\href
  {\doibase 10.1016/j.aop.2020.168360} {\bibfield  {journal} {\bibinfo
  {journal} {Annals Phys.}\ }\textbf {\bibinfo {volume} {424}},\ \bibinfo
  {pages} {168360} (\bibinfo {year} {2021})},\ \Eprint
  {http://arxiv.org/abs/2010.02254} {arXiv:2010.02254 [cond-mat.str-el]}
  \BibitemShut {NoStop}%
\bibitem [{\citenamefont {Hirono}\ \emph {et~al.}(2024)\citenamefont {Hirono},
  \citenamefont {You}, \citenamefont {Angus},\ and\ \citenamefont
  {Cho}}]{Hirono:2022dci}%
  \BibitemOpen
  \bibfield  {author} {\bibinfo {author} {\bibfnamefont {Y.}~\bibnamefont
  {Hirono}}, \bibinfo {author} {\bibfnamefont {M.}~\bibnamefont {You}},
  \bibinfo {author} {\bibfnamefont {S.}~\bibnamefont {Angus}}, \ and\ \bibinfo
  {author} {\bibfnamefont {G.~Y.}\ \bibnamefont {Cho}},\ }\href {\doibase
  10.21468/SciPostPhys.16.2.050} {\bibfield  {journal} {\bibinfo  {journal}
  {SciPost Phys.}\ }\textbf {\bibinfo {volume} {16}},\ \bibinfo {pages} {050}
  (\bibinfo {year} {2024})},\ \Eprint {http://arxiv.org/abs/2207.00854}
  {arXiv:2207.00854 [cond-mat.str-el]} \BibitemShut {NoStop}%
\bibitem [{\citenamefont {Iqbal}(2024)}]{Iqbal:2024pee}%
  \BibitemOpen
  \bibfield  {author} {\bibinfo {author} {\bibfnamefont {N.}~\bibnamefont
  {Iqbal}}\ }(\bibinfo {year} {2024})\ \Eprint
  {http://arxiv.org/abs/2407.20815} {arXiv:2407.20815 [hep-th]} \BibitemShut
  {NoStop}%
\bibitem [{\citenamefont {Das}\ \emph {et~al.}(2024)\citenamefont {Das},
  \citenamefont {Florio}, \citenamefont {Iqbal},\ and\ \citenamefont
  {Poovuttikul}}]{Das:2023nwl}%
  \BibitemOpen
  \bibfield  {author} {\bibinfo {author} {\bibfnamefont {A.}~\bibnamefont
  {Das}}, \bibinfo {author} {\bibfnamefont {A.}~\bibnamefont {Florio}},
  \bibinfo {author} {\bibfnamefont {N.}~\bibnamefont {Iqbal}}, \ and\ \bibinfo
  {author} {\bibfnamefont {N.}~\bibnamefont {Poovuttikul}},\ }\href {\doibase
  10.21468/SciPostPhys.17.3.085} {\bibfield  {journal} {\bibinfo  {journal}
  {SciPost Phys.}\ }\textbf {\bibinfo {volume} {17}},\ \bibinfo {pages} {085}
  (\bibinfo {year} {2024})},\ \Eprint {http://arxiv.org/abs/2309.14438}
  {arXiv:2309.14438 [hep-th]} \BibitemShut {NoStop}%
\bibitem [{\citenamefont {Berges}\ \emph {et~al.}(2009)\citenamefont {Berges},
  \citenamefont {Scheffler},\ and\ \citenamefont {Sexty}}]{Berges:2008mr}%
  \BibitemOpen
  \bibfield  {author} {\bibinfo {author} {\bibfnamefont {J.}~\bibnamefont
  {Berges}}, \bibinfo {author} {\bibfnamefont {S.}~\bibnamefont {Scheffler}}, \
  and\ \bibinfo {author} {\bibfnamefont {D.}~\bibnamefont {Sexty}},\ }\href
  {\doibase 10.1016/j.physletb.2009.10.032} {\bibfield  {journal} {\bibinfo
  {journal} {Phys. Lett. B}\ }\textbf {\bibinfo {volume} {681}},\ \bibinfo
  {pages} {362} (\bibinfo {year} {2009})},\ \Eprint
  {http://arxiv.org/abs/0811.4293} {arXiv:0811.4293 [hep-ph]} \BibitemShut
  {NoStop}%
\bibitem [{\citenamefont {Kraichnan}(1971)}]{kraichnan1971inertial}%
  \BibitemOpen
  \bibfield  {author} {\bibinfo {author} {\bibfnamefont {R.~H.}\ \bibnamefont
  {Kraichnan}},\ }\href {\doibase 10.1017/S0022112071001216} {\bibfield
  {journal} {\bibinfo  {journal} {J. Fluid Mech.}\ }\textbf {\bibinfo {volume}
  {47}},\ \bibinfo {pages} {525} (\bibinfo {year} {1971})}\BibitemShut
  {NoStop}%
\bibitem [{\citenamefont {Fj{\o}rtoft}(1953)}]{fjortoft1953changes}%
  \BibitemOpen
  \bibfield  {author} {\bibinfo {author} {\bibfnamefont {R.}~\bibnamefont
  {Fj{\o}rtoft}},\ }\href {\doibase 10.1111/j.2153-3490.1953.tb01051.x}
  {\bibfield  {journal} {\bibinfo  {journal} {Tellus}\ }\textbf {\bibinfo
  {volume} {5}},\ \bibinfo {pages} {225} (\bibinfo {year} {1953})}\BibitemShut
  {NoStop}%
\bibitem [{\citenamefont {Tabeling}(2002)}]{tabeling2002two}%
  \BibitemOpen
  \bibfield  {author} {\bibinfo {author} {\bibfnamefont {P.}~\bibnamefont
  {Tabeling}},\ }\href {\doibase 10.1016/S0370-1573(01)00064-3} {\bibfield
  {journal} {\bibinfo  {journal} {Phys. Rept.}\ }\textbf {\bibinfo {volume}
  {362}},\ \bibinfo {pages} {1} (\bibinfo {year} {2002})}\BibitemShut {NoStop}%
\bibitem [{\citenamefont {Alexakis}\ and\ \citenamefont
  {Biferale}(2018)}]{alexakis2018cascades}%
  \BibitemOpen
  \bibfield  {author} {\bibinfo {author} {\bibfnamefont {A.}~\bibnamefont
  {Alexakis}}\ and\ \bibinfo {author} {\bibfnamefont {L.}~\bibnamefont
  {Biferale}},\ }\href {\doibase 10.1016/j.physrep.2018.08.001} {\bibfield
  {journal} {\bibinfo  {journal} {Phys. Rept.}\ }\textbf {\bibinfo {volume}
  {767--769}},\ \bibinfo {pages} {1} (\bibinfo {year} {2018})}\BibitemShut
  {NoStop}%
\bibitem [{\citenamefont {Moffatt}(1969)}]{moffatt1969degree}%
  \BibitemOpen
  \bibfield  {author} {\bibinfo {author} {\bibfnamefont {H.~K.}\ \bibnamefont
  {Moffatt}},\ }\href {\doibase 10.1017/S0022112069000991} {\bibfield
  {journal} {\bibinfo  {journal} {J. Fluid Mech.}\ }\textbf {\bibinfo {volume}
  {35}},\ \bibinfo {pages} {117} (\bibinfo {year} {1969})}\BibitemShut
  {NoStop}%
\bibitem [{\citenamefont {Biferale}\ \emph {et~al.}(2012)\citenamefont
  {Biferale}, \citenamefont {Musacchio},\ and\ \citenamefont
  {Toschi}}]{biferale2012inverse}%
  \BibitemOpen
  \bibfield  {author} {\bibinfo {author} {\bibfnamefont {L.}~\bibnamefont
  {Biferale}}, \bibinfo {author} {\bibfnamefont {S.}~\bibnamefont {Musacchio}},
  \ and\ \bibinfo {author} {\bibfnamefont {F.}~\bibnamefont {Toschi}},\ }\href
  {\doibase 10.1103/PhysRevLett.108.164501} {\bibfield  {journal} {\bibinfo
  {journal} {Phys. Rev. Lett.}\ }\textbf {\bibinfo {volume} {108}},\ \bibinfo
  {pages} {164501} (\bibinfo {year} {2012})}\BibitemShut {NoStop}%
\bibitem [{\citenamefont {Sogabe}\ and\ \citenamefont
  {Yamamoto}(2019)}]{Sogabe:2019gif}%
  \BibitemOpen
  \bibfield  {author} {\bibinfo {author} {\bibfnamefont {N.}~\bibnamefont
  {Sogabe}}\ and\ \bibinfo {author} {\bibfnamefont {N.}~\bibnamefont
  {Yamamoto}},\ }\href {\doibase 10.1103/PhysRevD.99.125003} {\bibfield
  {journal} {\bibinfo  {journal} {Phys. Rev. D}\ }\textbf {\bibinfo {volume}
  {99}},\ \bibinfo {pages} {125003} (\bibinfo {year} {2019})},\ \Eprint
  {http://arxiv.org/abs/1903.02846} {arXiv:1903.02846 [hep-th]} \BibitemShut
  {NoStop}%
\bibitem [{\citenamefont {Hidaka}\ \emph {et~al.}(2020)\citenamefont {Hidaka},
  \citenamefont {Nitta},\ and\ \citenamefont {Yokokura}}]{Hidaka:2020iaz}%
  \BibitemOpen
  \bibfield  {author} {\bibinfo {author} {\bibfnamefont {Y.}~\bibnamefont
  {Hidaka}}, \bibinfo {author} {\bibfnamefont {M.}~\bibnamefont {Nitta}}, \
  and\ \bibinfo {author} {\bibfnamefont {R.}~\bibnamefont {Yokokura}},\ }\href
  {\doibase 10.1016/j.physletb.2020.135672} {\bibfield  {journal} {\bibinfo
  {journal} {Phys. Lett. B}\ }\textbf {\bibinfo {volume} {808}},\ \bibinfo
  {pages} {135672} (\bibinfo {year} {2020})},\ \Eprint
  {http://arxiv.org/abs/2006.12532} {arXiv:2006.12532 [hep-th]} \BibitemShut
  {NoStop}%
\bibitem [{\citenamefont {Hidaka}\ \emph
  {et~al.}(2021{\natexlab{b}})\citenamefont {Hidaka}, \citenamefont {Nitta},\
  and\ \citenamefont {Yokokura}}]{Hidaka:2020izy}%
  \BibitemOpen
  \bibfield  {author} {\bibinfo {author} {\bibfnamefont {Y.}~\bibnamefont
  {Hidaka}}, \bibinfo {author} {\bibfnamefont {M.}~\bibnamefont {Nitta}}, \
  and\ \bibinfo {author} {\bibfnamefont {R.}~\bibnamefont {Yokokura}},\ }\href
  {\doibase 10.1007/JHEP01(2021)173} {\bibfield  {journal} {\bibinfo  {journal}
  {JHEP}\ }\textbf {\bibinfo {volume} {01}},\ \bibinfo {pages} {173} (\bibinfo
  {year} {2021}{\natexlab{b}})},\ \Eprint {http://arxiv.org/abs/2009.14368}
  {arXiv:2009.14368 [hep-th]} \BibitemShut {NoStop}%
\bibitem [{\citenamefont {Choi}\ \emph {et~al.}(2023)\citenamefont {Choi},
  \citenamefont {Lam},\ and\ \citenamefont {Shao}}]{Choi:2022fgx}%
  \BibitemOpen
  \bibfield  {author} {\bibinfo {author} {\bibfnamefont {Y.}~\bibnamefont
  {Choi}}, \bibinfo {author} {\bibfnamefont {H.~T.}\ \bibnamefont {Lam}}, \
  and\ \bibinfo {author} {\bibfnamefont {S.-H.}\ \bibnamefont {Shao}},\ }\href
  {\doibase 10.1007/JHEP09(2023)067} {\bibfield  {journal} {\bibinfo  {journal}
  {JHEP}\ }\textbf {\bibinfo {volume} {09}},\ \bibinfo {pages} {067} (\bibinfo
  {year} {2023})},\ \Eprint {http://arxiv.org/abs/2212.04499} {arXiv:2212.04499
  [hep-th]} \BibitemShut {NoStop}%
\bibitem [{\citenamefont {Yokokura}(2022)}]{Yokokura:2022alv}%
  \BibitemOpen
  \bibfield  {author} {\bibinfo {author} {\bibfnamefont {R.}~\bibnamefont
  {Yokokura}},\ }\href@noop {} {\  (\bibinfo {year} {2022})},\ \Eprint
  {http://arxiv.org/abs/2212.05001} {arXiv:2212.05001 [hep-th]} \BibitemShut
  {NoStop}%
\bibitem [{\citenamefont {Yamamoto}\ and\ \citenamefont
  {Yokokura}(2023)}]{Yamamoto:2023uzq}%
  \BibitemOpen
  \bibfield  {author} {\bibinfo {author} {\bibfnamefont {N.}~\bibnamefont
  {Yamamoto}}\ and\ \bibinfo {author} {\bibfnamefont {R.}~\bibnamefont
  {Yokokura}},\ }\href {\doibase 10.1007/JHEP07(2023)045} {\bibfield  {journal}
  {\bibinfo  {journal} {JHEP}\ }\textbf {\bibinfo {volume} {07}},\ \bibinfo
  {pages} {045} (\bibinfo {year} {2023})},\ \Eprint
  {http://arxiv.org/abs/2305.01234} {arXiv:2305.01234 [hep-th]} \BibitemShut
  {NoStop}%
\bibitem [{\citenamefont {Bergman}\ \emph {et~al.}(2011)\citenamefont
  {Bergman}, \citenamefont {Jokela}, \citenamefont {Lifschytz},\ and\
  \citenamefont {Lippert}}]{Bergman:2011rf}%
  \BibitemOpen
  \bibfield  {author} {\bibinfo {author} {\bibfnamefont {O.}~\bibnamefont
  {Bergman}}, \bibinfo {author} {\bibfnamefont {N.}~\bibnamefont {Jokela}},
  \bibinfo {author} {\bibfnamefont {G.}~\bibnamefont {Lifschytz}}, \ and\
  \bibinfo {author} {\bibfnamefont {M.}~\bibnamefont {Lippert}},\ }\href
  {\doibase 10.1007/JHEP10(2011)034} {\bibfield  {journal} {\bibinfo  {journal}
  {JHEP}\ }\textbf {\bibinfo {volume} {10}},\ \bibinfo {pages} {034} (\bibinfo
  {year} {2011})},\ \Eprint {http://arxiv.org/abs/1106.3883} {arXiv:1106.3883
  [hep-th]} \BibitemShut {NoStop}%
\bibitem [{\citenamefont {Ooguri}\ and\ \citenamefont
  {Oshikawa}(2012)}]{Ooguri:2011aa}%
  \BibitemOpen
  \bibfield  {author} {\bibinfo {author} {\bibfnamefont {H.}~\bibnamefont
  {Ooguri}}\ and\ \bibinfo {author} {\bibfnamefont {M.}~\bibnamefont
  {Oshikawa}},\ }\href {\doibase 10.1103/PhysRevLett.108.161803} {\bibfield
  {journal} {\bibinfo  {journal} {Phys. Rev. Lett.}\ }\textbf {\bibinfo
  {volume} {108}},\ \bibinfo {pages} {161803} (\bibinfo {year} {2012})},\
  \Eprint {http://arxiv.org/abs/1112.1414} {arXiv:1112.1414
  [cond-mat.mes-hall]} \BibitemShut {NoStop}%
\bibitem [{\citenamefont {Yamamoto}\ and\ \citenamefont
  {Yokokura}(2022)}]{Yamamoto:2022vrh}%
  \BibitemOpen
  \bibfield  {author} {\bibinfo {author} {\bibfnamefont {N.}~\bibnamefont
  {Yamamoto}}\ and\ \bibinfo {author} {\bibfnamefont {R.}~\bibnamefont
  {Yokokura}},\ }\href {\doibase 10.1103/PhysRevD.106.105004} {\bibfield
  {journal} {\bibinfo  {journal} {Phys. Rev. D}\ }\textbf {\bibinfo {volume}
  {106}},\ \bibinfo {pages} {105004} (\bibinfo {year} {2022})},\ \Eprint
  {http://arxiv.org/abs/2203.02727} {arXiv:2203.02727 [hep-th]} \BibitemShut
  {NoStop}%
\bibitem [{\citenamefont {Choi}\ \emph {et~al.}(2022)\citenamefont {Choi},
  \citenamefont {Lam},\ and\ \citenamefont {Shao}}]{Choi:2022jqy}%
  \BibitemOpen
  \bibfield  {author} {\bibinfo {author} {\bibfnamefont {Y.}~\bibnamefont
  {Choi}}, \bibinfo {author} {\bibfnamefont {H.~T.}\ \bibnamefont {Lam}}, \
  and\ \bibinfo {author} {\bibfnamefont {S.-H.}\ \bibnamefont {Shao}},\ }\href
  {\doibase 10.1103/PhysRevLett.129.161601} {\bibfield  {journal} {\bibinfo
  {journal} {Phys. Rev. Lett.}\ }\textbf {\bibinfo {volume} {129}},\ \bibinfo
  {pages} {161601} (\bibinfo {year} {2022})},\ \Eprint
  {http://arxiv.org/abs/2205.05086} {arXiv:2205.05086 [hep-th]} \BibitemShut
  {NoStop}%
\bibitem [{\citenamefont {Cordova}\ and\ \citenamefont
  {Ohmori}(2023)}]{Cordova:2022ieu}%
  \BibitemOpen
  \bibfield  {author} {\bibinfo {author} {\bibfnamefont {C.}~\bibnamefont
  {Cordova}}\ and\ \bibinfo {author} {\bibfnamefont {K.}~\bibnamefont
  {Ohmori}},\ }\href {\doibase 10.1103/PhysRevX.13.011034} {\bibfield
  {journal} {\bibinfo  {journal} {Phys. Rev. X}\ }\textbf {\bibinfo {volume}
  {13}},\ \bibinfo {pages} {011034} (\bibinfo {year} {2023})},\ \Eprint
  {http://arxiv.org/abs/2205.06243} {arXiv:2205.06243 [hep-th]} \BibitemShut
  {NoStop}%
\bibitem [{\citenamefont {Joyce}\ and\ \citenamefont
  {Shaposhnikov}(1997)}]{Joyce:1997uy}%
  \BibitemOpen
  \bibfield  {author} {\bibinfo {author} {\bibfnamefont {M.}~\bibnamefont
  {Joyce}}\ and\ \bibinfo {author} {\bibfnamefont {M.~E.}\ \bibnamefont
  {Shaposhnikov}},\ }\href {\doibase 10.1103/PhysRevLett.79.1193} {\bibfield
  {journal} {\bibinfo  {journal} {Phys. Rev. Lett.}\ }\textbf {\bibinfo
  {volume} {79}},\ \bibinfo {pages} {1193} (\bibinfo {year} {1997})},\ \Eprint
  {http://arxiv.org/abs/astro-ph/9703005} {arXiv:astro-ph/9703005} \BibitemShut
  {NoStop}%
\bibitem [{\citenamefont {Akamatsu}\ and\ \citenamefont
  {Yamamoto}(2013)}]{Akamatsu:2013pjd}%
  \BibitemOpen
  \bibfield  {author} {\bibinfo {author} {\bibfnamefont {Y.}~\bibnamefont
  {Akamatsu}}\ and\ \bibinfo {author} {\bibfnamefont {N.}~\bibnamefont
  {Yamamoto}},\ }\href {\doibase 10.1103/PhysRevLett.111.052002} {\bibfield
  {journal} {\bibinfo  {journal} {Phys. Rev. Lett.}\ }\textbf {\bibinfo
  {volume} {111}},\ \bibinfo {pages} {052002} (\bibinfo {year} {2013})},\
  \Eprint {http://arxiv.org/abs/1302.2125} {arXiv:1302.2125 [nucl-th]}
  \BibitemShut {NoStop}%
\bibitem [{\citenamefont {Crossley}\ \emph {et~al.}(2017)\citenamefont
  {Crossley}, \citenamefont {Glorioso},\ and\ \citenamefont
  {Liu}}]{Crossley:2015evo}%
  \BibitemOpen
  \bibfield  {author} {\bibinfo {author} {\bibfnamefont {M.}~\bibnamefont
  {Crossley}}, \bibinfo {author} {\bibfnamefont {P.}~\bibnamefont {Glorioso}},
  \ and\ \bibinfo {author} {\bibfnamefont {H.}~\bibnamefont {Liu}},\ }\href
  {\doibase 10.1007/JHEP09(2017)095} {\bibfield  {journal} {\bibinfo  {journal}
  {JHEP}\ }\textbf {\bibinfo {volume} {09}},\ \bibinfo {pages} {095} (\bibinfo
  {year} {2017})},\ \Eprint {http://arxiv.org/abs/1511.03646} {arXiv:1511.03646
  [hep-th]} \BibitemShut {NoStop}%
\bibitem [{\citenamefont {Liu}\ and\ \citenamefont
  {Glorioso}(2018)}]{Liu:2018kfw}%
  \BibitemOpen
  \bibfield  {author} {\bibinfo {author} {\bibfnamefont {H.}~\bibnamefont
  {Liu}}\ and\ \bibinfo {author} {\bibfnamefont {P.}~\bibnamefont {Glorioso}},\
  }\href {\doibase 10.22323/1.305.0008} {\bibfield  {journal} {\bibinfo
  {journal} {PoS}\ }\textbf {\bibinfo {volume} {TASI2017}},\ \bibinfo {pages}
  {008} (\bibinfo {year} {2018})},\ \Eprint {http://arxiv.org/abs/1805.09331}
  {arXiv:1805.09331 [hep-th]} \BibitemShut {NoStop}%
\bibitem [{\citenamefont {Hirono}\ \emph {et~al.}(2015)\citenamefont {Hirono},
  \citenamefont {Kharzeev},\ and\ \citenamefont {Yin}}]{Hirono:2015rla}%
  \BibitemOpen
  \bibfield  {author} {\bibinfo {author} {\bibfnamefont {Y.}~\bibnamefont
  {Hirono}}, \bibinfo {author} {\bibfnamefont {D.}~\bibnamefont {Kharzeev}}, \
  and\ \bibinfo {author} {\bibfnamefont {Y.}~\bibnamefont {Yin}},\ }\href
  {\doibase 10.1103/PhysRevD.92.125031} {\bibfield  {journal} {\bibinfo
  {journal} {Phys. Rev.}\ }\textbf {\bibinfo {volume} {D92}},\ \bibinfo {pages}
  {125031} (\bibinfo {year} {2015})},\ \Eprint
  {http://arxiv.org/abs/1509.07790} {arXiv:1509.07790 [hep-th]} \BibitemShut
  {NoStop}%
\bibitem [{\citenamefont {Kraichnan}(1967)}]{kraichnan1967}%
  \BibitemOpen
  \bibfield  {author} {\bibinfo {author} {\bibfnamefont {R.~H.}\ \bibnamefont
  {Kraichnan}},\ }\href {\doibase 10.1063/1.1762301} {\bibfield  {journal}
  {\bibinfo  {journal} {Phys. Fluids}\ }\textbf {\bibinfo {volume} {10}},\
  \bibinfo {pages} {1417} (\bibinfo {year} {1967})}\BibitemShut {NoStop}%
\bibitem [{\citenamefont {Boffetta}\ and\ \citenamefont
  {Ecke}(2012)}]{boffetta2012two}%
  \BibitemOpen
  \bibfield  {author} {\bibinfo {author} {\bibfnamefont {G.}~\bibnamefont
  {Boffetta}}\ and\ \bibinfo {author} {\bibfnamefont {R.~E.}\ \bibnamefont
  {Ecke}},\ }\href {\doibase 10.1146/annurev-fluid-120710-101240} {\bibfield
  {journal} {\bibinfo  {journal} {Annu. Rev. Fluid Mech.}\ }\textbf {\bibinfo
  {volume} {44}},\ \bibinfo {pages} {427} (\bibinfo {year} {2012})}\BibitemShut
  {NoStop}%
\bibitem [{\citenamefont {Tashiro}\ \emph {et~al.}(2012)\citenamefont
  {Tashiro}, \citenamefont {Vachaspati},\ and\ \citenamefont
  {Vilenkin}}]{Tashiro:2012mf}%
  \BibitemOpen
  \bibfield  {author} {\bibinfo {author} {\bibfnamefont {H.}~\bibnamefont
  {Tashiro}}, \bibinfo {author} {\bibfnamefont {T.}~\bibnamefont {Vachaspati}},
  \ and\ \bibinfo {author} {\bibfnamefont {A.}~\bibnamefont {Vilenkin}},\
  }\href {\doibase 10.1103/PhysRevD.86.105033} {\bibfield  {journal} {\bibinfo
  {journal} {Phys. Rev. D}\ }\textbf {\bibinfo {volume} {86}},\ \bibinfo
  {pages} {105033} (\bibinfo {year} {2012})},\ \Eprint
  {http://arxiv.org/abs/1206.5549} {arXiv:1206.5549 [astro-ph.CO]} \BibitemShut
  {NoStop}%
\bibitem [{\citenamefont {Yamamoto}(2016)}]{Yamamoto:2016xtu}%
  \BibitemOpen
  \bibfield  {author} {\bibinfo {author} {\bibfnamefont {N.}~\bibnamefont
  {Yamamoto}},\ }\href {\doibase 10.1103/PhysRevD.93.125016} {\bibfield
  {journal} {\bibinfo  {journal} {Phys. Rev. D}\ }\textbf {\bibinfo {volume}
  {93}},\ \bibinfo {pages} {125016} (\bibinfo {year} {2016})},\ \Eprint
  {http://arxiv.org/abs/1603.08864} {arXiv:1603.08864 [hep-th]} \BibitemShut
  {NoStop}%
\bibitem [{\citenamefont {Buividovich}\ and\ \citenamefont
  {Ulybyshev}(2016)}]{Buividovich:2015jfa}%
  \BibitemOpen
  \bibfield  {author} {\bibinfo {author} {\bibfnamefont {P.~V.}\ \bibnamefont
  {Buividovich}}\ and\ \bibinfo {author} {\bibfnamefont {M.~V.}\ \bibnamefont
  {Ulybyshev}},\ }\href {\doibase 10.1103/PhysRevD.94.025009} {\bibfield
  {journal} {\bibinfo  {journal} {Phys. Rev. D}\ }\textbf {\bibinfo {volume}
  {94}},\ \bibinfo {pages} {025009} (\bibinfo {year} {2016})},\ \Eprint
  {http://arxiv.org/abs/1509.02076} {arXiv:1509.02076 [hep-th]} \BibitemShut
  {NoStop}%
\bibitem [{\citenamefont {Mace}\ \emph {et~al.}(2020)\citenamefont {Mace},
  \citenamefont {Mueller}, \citenamefont {Schlichting},\ and\ \citenamefont
  {Sharma}}]{Mace:2019cqo}%
  \BibitemOpen
  \bibfield  {author} {\bibinfo {author} {\bibfnamefont {M.}~\bibnamefont
  {Mace}}, \bibinfo {author} {\bibfnamefont {N.}~\bibnamefont {Mueller}},
  \bibinfo {author} {\bibfnamefont {S.}~\bibnamefont {Schlichting}}, \ and\
  \bibinfo {author} {\bibfnamefont {S.}~\bibnamefont {Sharma}},\ }\href
  {\doibase 10.1103/PhysRevLett.124.191604} {\bibfield  {journal} {\bibinfo
  {journal} {Phys. Rev. Lett.}\ }\textbf {\bibinfo {volume} {124}},\ \bibinfo
  {pages} {191604} (\bibinfo {year} {2020})},\ \Eprint
  {http://arxiv.org/abs/1910.01654} {arXiv:1910.01654 [hep-ph]} \BibitemShut
  {NoStop}%
\bibitem [{\citenamefont {Turner}\ and\ \citenamefont
  {Widrow}(1988)}]{Turner:1987bw}%
  \BibitemOpen
  \bibfield  {author} {\bibinfo {author} {\bibfnamefont {M.~S.}\ \bibnamefont
  {Turner}}\ and\ \bibinfo {author} {\bibfnamefont {L.~M.}\ \bibnamefont
  {Widrow}},\ }\href {\doibase 10.1103/PhysRevD.37.2743} {\bibfield  {journal}
  {\bibinfo  {journal} {Phys. Rev. D}\ }\textbf {\bibinfo {volume} {37}},\
  \bibinfo {pages} {2743} (\bibinfo {year} {1988})}\BibitemShut {NoStop}%
\bibitem [{\citenamefont {Ratra}(1992)}]{Ratra:1991bn}%
  \BibitemOpen
  \bibfield  {author} {\bibinfo {author} {\bibfnamefont {B.}~\bibnamefont
  {Ratra}},\ }\href {\doibase 10.1086/186384} {\bibfield  {journal} {\bibinfo
  {journal} {Astrophys. J. Lett.}\ }\textbf {\bibinfo {volume} {391}},\
  \bibinfo {pages} {L1} (\bibinfo {year} {1992})}\BibitemShut {NoStop}%
\end{thebibliography}%

\end{document}